\documentclass[twocolumn,showpacs,preprintnumbers,amsmath,amssymb,prb]{revtex4}

\usepackage{graphicx}
\usepackage{dcolumn}
\usepackage{bm}
\usepackage{color}

\begin{document}

\preprint{}

\title{Observation of elastic anomalies driven by coexisting dynamical spin Jahn-Teller effect and dynamical molecular spin state in paramagnetic phase of the frustrated MgCr$_2$O$_4$}

\author{Tadataka Watanabe$^1$}
\author{Shin-Ichi Ishikawa$^1$}
\author{Haruki Suzuki$^1$}
\author{Yusuke Kousaka$^2$}
\author{Keisuke Tomiyasu$^3$}
\affiliation{$^1$Department of Physics, College of Science and Technology (CST), Nihon University, Chiyoda, Tokyo 101-8308, Japan}
\affiliation{$^2$Department of Physics and Mathematics, Aoyama-Gakuin University, Sagamihara, Kanagawa 229-8558, Japan}
\affiliation{$^3$Department of Physics, Tohoku University, Aoba, Sendai 980-8578, Japan}
\date{\today}

\begin{abstract}
Ultrasound velocity measurements of magnesium chromite spinel MgCr$_2$O$_4$ reveal elastic anomalies in the paramagnetic phase that are characterized as due to geometrical frustration. The temperature dependence of the tetragonal shear modulus $(C_{11}-C_{12})/2$ exhibits huge Curie-type softening, which should be the precursor to spin Jahn-Teller distortion in the antiferromagnetic phase. The trigonal shear modulus $C_{44}$ exhibits nonmonotonic temperature dependence with a characteristic minimum at $\sim$50 K, indicating a coupling of the lattice to dynamical molecular spin state. These results strongly suggest the coexistence of dynamical spin Jahn-Teller effect and dynamical molecular spin state in the paramagnetic phase, which is compatible with the coexistence of magnetostructural order and dynamical molecular spin state in the antiferromagnetic phase.
\end{abstract}


\maketitle

\section{Introduction}
Geometrically frustrated magnets have attracted considerable interest because the vast degeneracy of their ground states gives rise to a rich variety of novel phenomena \cite{Moessner}. One of the most highly frustrated spin systems is the nearest-neighbor Heisenberg antiferromagnet on the pyrochlore lattice, where the magnetic sites form a network of corner-sharing tetrahedra. The prototypical examples for this highly frustrated system are chromite spinels $A$Cr$_2$O$_4$ with non-magnetic ions $A$ = Mg, Zn, and Cd. The magnetic properties of these compounds are fully dominated by the Jahn-Teller (JT)-inactive Cr$^{3+}$ with spin $S$ = 3/2 residing on the pyrochlore network. The high values of frustration parameter $f=|\Theta_W/T_N|$, the ratio of Weiss and N$\acute{e}$el temperatures, indicate the presence of strong geometrical frustration: $f\simeq$ 30 with $\Theta_W$ = -370 K and $T_N$ = 12.5 K for MgCr$_2$O$_4$, $f\simeq$ 33 with $\Theta_W$ = -390 K and $T_N$ = 12.0 K for ZnCr$_2$O$_4$, and $f\simeq$ 9 with $\Theta_W$ = -70 K and $T_N$ = 7.8 K for CdCr$_2$O$_4$ \cite{Ueda}.

In $A$Cr$_2$O$_4$, the antiferromagnetic (AF) transition at $T_N$ coincides with a tetragonal lattice distortion with a compression along the $c$-axis for MgCr$_2$O$_4$ and ZnCr$_2$O$_4$ \cite{Lee1,Ortega-San-Martin}, and an elongation for CdCr$_2$O$_4$ \cite{Chung}. This magnetostructural ordering is explained by the spin JT mechanism via the magnetoelastic coupling, where local distortions of the tetrahedra release the frustration in the nearest-neighbor AF interactions \cite{Yamashita,Tchernyshyov1}. The very small magnitudes of the tetragonal lattice strain of the order 10$^{-3}$ imply the presence of a nonuniform component in the lattice distortion, where the neighboring tetrahedra are alternately stretched and compressed along the same crystal axis \cite{Lee1,Ortega-San-Martin,Chung,Tchernyshyov2}. Indeed, a recent study of ZnCr$_2$O$_4$ using synchrotron X-ray and neutron scattering experiments revealed complex spin-lattice order with the nonuniform lattice distortion \cite{Ji}.

In the paramagnetic (PM) phase of ZnCr$_2$O$_4$ and MgCr$_2$O$_4$, inelastic neutron scattering (INS) experiments observed quasielastic magnetic scattering, indicating the presence of strong spin fluctuations due to frustration \cite{Lee1,Lee2,Suzuki,Tomiyasu}. Interestingly, this quasigapless mode was characterized as fluctuations of AF hexagonal spin clusters (AF hexamers) in the pyrochlore lattice \cite{Lee2,Tomiyasu}. It is natural to expect that this dynamical molecular spin state vanishes in the AF phase as a consequence of the release of frustration. Actually, however, the INS studies revealed the appearance of gapped molecular spin excitations in the AF phase, which indicates the coexistence of the magnetostructural order and the zero point motion-like spin molecules \cite{Tomiyasu}.

The appearance of the magnetostructural order in $A$Cr$_2$O$_4$ means that the spin-lattice coupling plays a significant role in releasing the frustration; thus, the phonon spectra are expected to be affected by the frustration. Indeed, infrared and Raman spectroscopies in ZnCr$_2$O$_4$ and CdCr$_2$O$_4$ revealed that the optical phonon spectra exhibit anomalies of softening on cooling in the PM phase, and splitting in the AF phase \cite{Sushkov,Kant2,Aguilar}. In this paper, we present acoustic phonon information provided by ultrasound velocity measurements in magnesium chromite spinel MgCr$_2$O$_4$. The modified sound dispersions by magnetoelastic coupling allow one to extract detailed information about the interplay between spin and lattice degrees of freedom in a lower energy region near to the ground states. Thus ultrasound velocity measurements can be a useful tool in studying geometrically frustrated magnets \cite{Kino,Bhattacharjee,Watanabe1,Watanabe2}. We find elastic-mode-dependent sound-velocity anomalies in MgCr$_2$O$_4$ in the PM phase, similar to those observed in the past in ZnCr$_2$O$_4$ \cite{Kino}. The origins of these elastic anomalies are discussed in terms of frustration inherent in $A$Cr$_2$O$_4$.

\section{Experimental}
Ultrasound velocities were measured in single crystals of MgCr$_2$O$_4$ prepared by the floating zone method, where the phase comparison technique was used with longitudinal and transverse sound waves at a frequency of 30 MHz. The ultrasounds were generated and detected by LiNbO$_3$ transducers glued on the parallel mirror surfaces of the crystal. We measured sound velocities in all the symmetrically independent elastic moduli in the cubic crystal: compression modulus $C_{11}$, tetragonal shear modulus $\frac{C_{11}-C_{12}}{2}\equiv C_t$, and trigonal shear modulus $C_{44}$. The respective measurements of $C_{11}$, $C_t$, and $C_{44}$ were performed by using longitudinal wave with propagation {\bf k}$\parallel$[100] and polarization {\bf u}$\parallel$[100], transverse wave with {\bf k}$\parallel$[110] and {\bf u}$\parallel$[1$\bar{1}$0], and transverse wave with {\bf k}$\parallel$[110] and {\bf u}$\parallel$[001].

\section{Results}
Figures 1(a)-(c) depict the longitudinal sound velocity $v_L$ in $C_{11}$ and the transverse sound velocity $v_T$ in $C_t$ and $C_{44}$ as functions of temperature ($T$), respectively. All the elastic moduli exhibit a jump at $T_N$ = 13 K. And, more importantly, all the elastic moduli exhibit elastic-mode-dependent anomalous $T$-dependence in the PM phase, from room temperature (300 K) down to $T_N$, that is different from the $T$-dependence usually observed in solids, namely monotonic hardening with decreasing $T$ \cite{Varshni}. These elastic anomalies should have magnetic origins where the spin degrees of freedom play significant roles but also taking into account the absence of orbital degeneracy in the $B$-site Cr$^{3+}$ in MgCr$_2$O$_4$. Such elastic anomalies are attributed to magnetoelastic coupling acting on the exchange interactions. In this mechanism, the exchange striction arises from a modulation of the exchange interactions by ultrasound as follows: \cite{Luthi}

\begin{equation}
H_{exs} = \sum_{ij}[J(\delta + u_i - u_j) - J(\delta)]S_i \cdot S_j.
\label{eq:ES1}
\end{equation}
Here $\delta = R_i-R_j$ is the distance between two magnetic ions, and $u_i$ is the displacement vector for the ion $R_i$. When a sound wave with polarization $u$ and propagation $k$ is given by $u$ = $u_0$exp[i($k \cdot r-\omega t$)], where $u_0$ and $\omega$ are the respective amplitude and frequency, the exchange striction of Eq.~(\ref{eq:ES1}) is rewritten as: \cite{Stern}

\begin{equation}
H_{exs} = \sum_{i}(\frac{\partial J}{\partial \delta} \cdot u_0)(k \cdot \delta)(S_i \cdot S_{i+\delta})e^{i(k \cdot R_i-\omega t)}.
\label{eq:ES2}
\end{equation}
Here the exponential is expanded to first order because with a 30-MHz ultrasound frequency $k\delta\ll1$. Eq.~(\ref{eq:ES2}) implies that both the longitudinal and transverse sound waves can couple to the spin system via the exchange striction mechanism, which depends on the directions of polarization $u$ and propagation $k$ relative to the exchange path $\delta$. Concerning the elastic anomalies in MgCr$_2$O$_4$, we emphasize that the anomalous $T$ variations of the elastic moduli in the PM phase are divided into two types: Curie-type -1/$T$ softening for $C_{11}(T)$ and $C_t(T)$ and nonmonotonic $T$-dependence with a characteristic minimum at $\sim$50 K for $C_{44}(T)$. Together these types imply that the elastic anomalies have different origins, which we shall discuss below.

\begin{figure}[t]
\begin{center}
\includegraphics[scale=0.4]{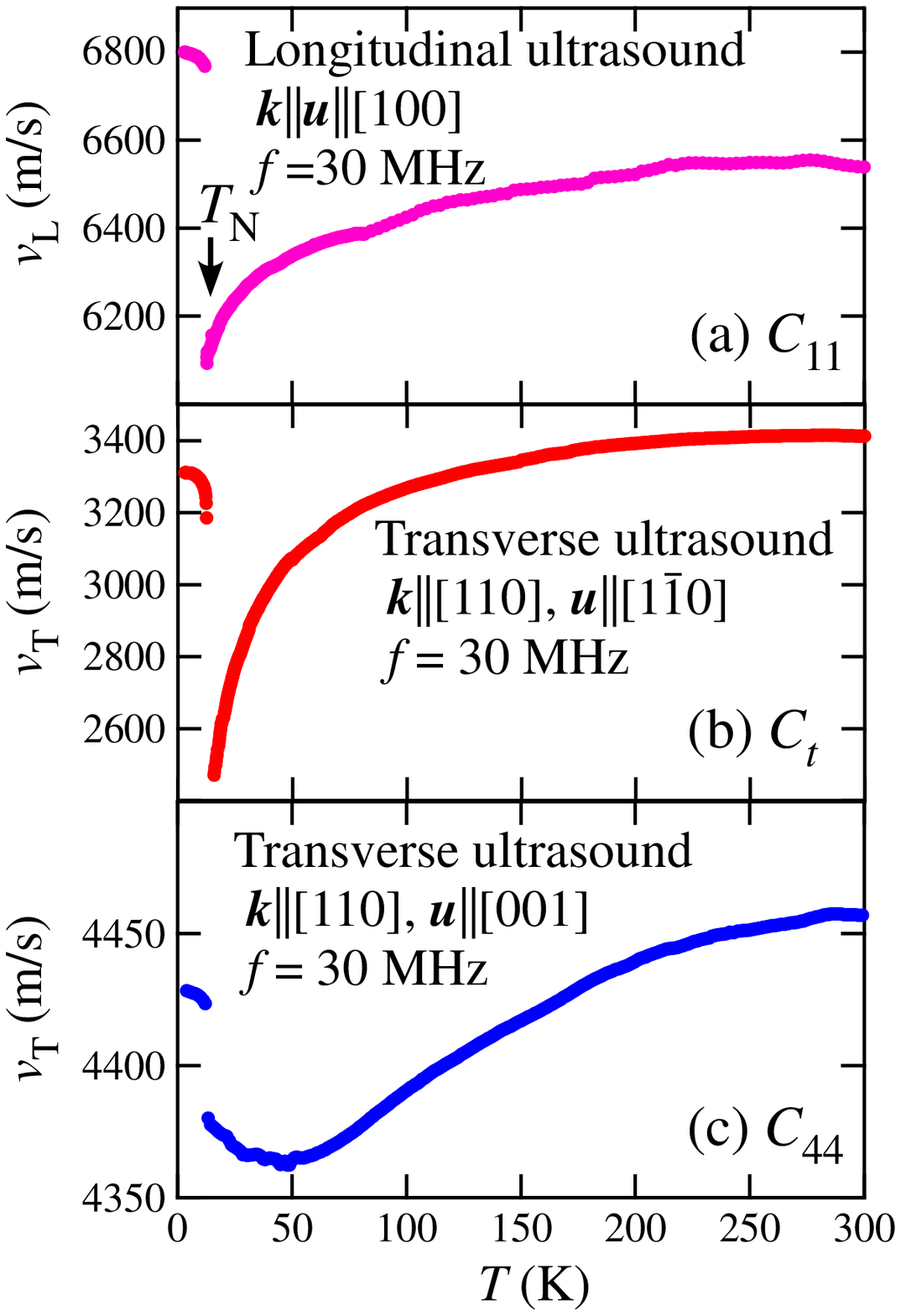}
\caption{\label{fig:Fig1} (Color online). Temperature dependence of the longitudinal ($v_L$) and transverse ($v_T$) sound velocities. (a) $C_{11}$, (b) $\frac{(C_{11}-C_{12})}{2}\equiv C_t$, and (c) $C_{44}$.}
\end{center}
\end{figure}

\section{Discussion}
\subsection{Curie-type softening in PM phase}
First we discuss the origin of the Curie-type softening in $C_{11}(T)$ and $C_t(T)$ in the PM phase. Taking into account that the relative change in softening is much larger in $C_t(T)$ ($\frac{\Delta C_t}{C_t}\sim$50$\%$) than in $C_{11}(T)$ ($\frac{\Delta C_{11}}{C_{11}}\sim$10$\%$), and that the compression modulus $C_{11}$ is written as $C_{11}= C_B+\frac{4}{3}C_t$ with $C_B$ the bulk modulus, the Curie-type softening should originate from a softening in the tetragonal shear modulus $C_t(T)$. Thus the Curie-type softening should be a precursor to the tetragonal structural transition at $T_N$. Such a precursor softening to the structural transition is well known to occur as a result of the Jahn-Teller (JT) effect in orbital-degenerate systems \cite{Luthi}. However, the absence of orbital degeneracy for the $B$-site Cr$^{3+}$ in MgCr$_2$O$_4$ rules out such orbital effects. Thus {\it the Curie-type softening is most probably due to a spin JT effect \cite{Yamashita,Tchernyshyov1}, which is compatible with the coincidence of AF ordering and tetragonal lattice compression at $T_N$}. The spin JT distortion is analogous to orbital distortion in the sense that a degenerate electronic system lowers its energy by lifting the degeneracy under structural distortion. In $A$Cr$_2$O$_4$, the spin JT distortion lifts the spin degeneracy (frustration) inherent in the pyrochlore lattice by modulating the balance of exchange interactions via the exchange striction mechanism expressed as Eq.~(\ref{eq:ES1}).

For the orbital JT effect, a JT-active ion and its set of ligands are considered to be a structural unit, where the crystal-field striction mechanism leads to the orbital JT distortion. $T$ dependence of elastic modulus $C_{\Gamma}(T)$ in the orbital JT systems is explained assuming the coupling of ultrasound to JT-active ions via the crystal-field striction mechanism, and the presence of inter-JT-active-ion interactions. A mean-field expression of $C_{\Gamma}(T)$ in the orbital JT systems is given as: \cite{Kino2, Hazama, Kataoka}

\begin{equation}
C_{\Gamma}(T)=C_{\Gamma}^0-g_{\Gamma}^{2} N \frac{\chi_{\Gamma}(T)}{(1-g^{'}_{\Gamma}\chi_{\Gamma}(T))},
\label{eq:SS}
\end{equation}
with $C^0_{\Gamma}$ the elastic constant without the orbital JT effect, $g_{\Gamma}$ the coupling constant of crystal field strain, $N$ the number of JT-active ions in the unit volume, $g^{'}_{\Gamma}$ the inter-JT-active-ion interaction, and $\chi_{\Gamma}(T)$ the strain susceptibility of a single JT-active ion. Eq.~(\ref{eq:SS}) expresses $T$ dependence of the strain susceptibility which is the response function to an applied strain, and its form is analogous to the mean-field expression of the magnetic susceptibility which is the response function of the magnetization to an applied magnetic field. In the orbital JT systems, the degenerate ground state is considered to couple strongly and selectively to the elastic modulus $C_{\Gamma}$ which has the same symmetry as the orbital JT distortion. For such a JT-active elastic mode, the strain susceptibility is dominated by Curie term $\chi_{\Gamma}(T)\sim1/T$ at low temperatures. Then Eq.~(\ref{eq:SS}) is rewritten as: \cite{Kino2, Hazama, Kataoka}

\begin{equation}
C_{\Gamma}(T) = C^0_{\Gamma} \frac{T-T_c}{T-{\it \theta}},
\label{eq:JT}
\end{equation}
with $\theta$ the inter-JT-active-ion interaction, and $T_c$ the second-order structural transition temperature. Here $\theta$ is positive/negative when the interaction is ferrodistortive/antiferrodistortive. Although the background $C^0_{\Gamma}$ in Eq.~(\ref{eq:JT}) generally exhibits a hardening with decreasing $T$ \cite{Varshni}, we here assume $C^0_{\Gamma}$ is constant because its hardening is negligibly small compared with the huge Curie-type softening in $C_{\Gamma}(T)$. In contrast to the JT-active soft mode $C_{\Gamma}$ expressed as Eq.~(\ref{eq:JT}), the elastic modulus $C_{\Gamma'}$ with the different symmetry from the orbital JT distortion is not affected by the degenerate ground state; $C_{\Gamma'}(T)$ exhibits the absence of anomaly, $C_{\Gamma'}(T)=C_{\Gamma'}^0$, originating from $\chi_{\Gamma'}=0$ in Eq.~(\ref{eq:SS}).

In analogy to the orbital JT effect, the Curie-type softening in $C_t(T)$ in MgCr$_2$O$_4$ can be explained in the spin-JT-effect-scenario assuming the coupling of a Cr$^{3+}$ structural unit to the external strain of ultrasound via the exchange striction mechanism expressed as Eqs.~(\ref{eq:ES1}) and (\ref{eq:ES2}), and the presence of inter-structural-unit interactions. Although the structural unit of the spin JT effect has not been identified so far, the mean-field expression of the soft mode $C_{\Gamma}(T)$ should have the same form as Eq.~(\ref{eq:JT}) with $C^0_{\Gamma}$ the elastic constant without the spin JT effect, ${\it \theta}$ the inter-structural-unit interaction, and $T_c$ the second-order structural transition temperature. For MgCr$_2$O$_4$, it is expected that ultrasound in $C_t$ generates the tetragonal exchange striction which promotes the tetragonal spin-lattice correlation, while the trigonal exchange striction is generated in $C_{44}$. Thus, since the spin JT distortion in MgCr$_2$O$_4$ has the tetragonal symmetry \cite{Ortega-San-Martin}, the tetragonal shear modulus $C_t(T)$ should exhibit Curie-type softening as expressed in Eq.~(\ref{eq:JT}), whereas its absence in the trigonal shear modulus is expressed by $C_{44}(T)=C_{44}^0$ in analogy to the orbital JT effect.

\begin{figure}[t]
\begin{center}
\includegraphics[scale=0.4]{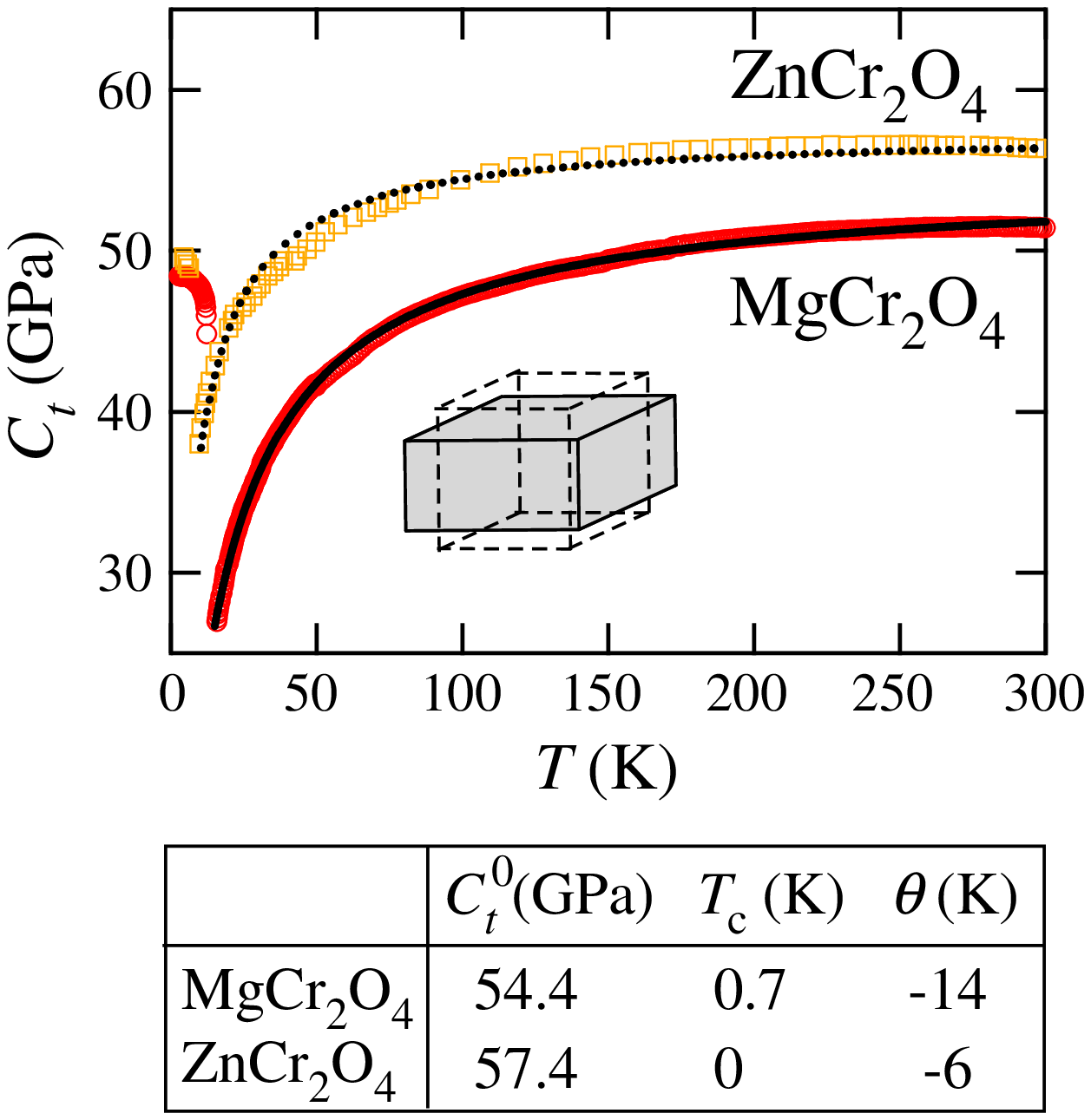}
\caption{\label{fig:Fig2} (Color online). The tetragonal shear modulus $C_t$ in MgCr$_2$O$_4$ (open circles, from Fig. 1(b)) and ZnCr$_2$O$_4$ (open squares, from Ref. [\cite{Kino}]) as functions of temperature. The solid and dotted curves are fits of the experimental data for MgCr$_2$O$_4$ and ZnCr$_2$O$_4$ to Eq.~(\ref{eq:JT}), respectively. Values for the fitting parameters are listed in the table beneath the figure. The inset shows a schematic of the tetragonal lattice deformation in $C_t$.}
\end{center}
\end{figure}

It is noted that the Curie-type softening in the tetragonal shear modulus $C_t(T)$ is observed in not only MgCr$_2$O$_4$ but also another highly frustrated spinel ZnCr$_2$O$_4$ \cite{Kino}. Figure 2 compares the tetragonal shear modulus $C_t$ in MgCr$_2$O$_4$ (open circles, from Fig. 1(b)) with that in ZnCr$_2$O$_4$ (open squares, from Ref. [\cite{Kino}]) as functions of $T$; fits of the experimental data for MgCr$_2$O$_4$ and ZnCr$_2$O$_4$ to Eq.~(\ref{eq:JT}) are also plotted as solid and dotted curves, respectively. The plots reproduce the experimental data given the fitted values of $C^0_t$, $T_c$, and ${\it \theta}$ listed in the table beneath the figure. For MgCr$_2$O$_4$ and ZnCr$_2$O$_4$, the respective fitted values of $T_c$ = 0.7 K and 0 K are lower than the experimentally-observed N$\acute{e}$el temperatures of $T_N$ = 13 K and 12.5 K, $T_c<T_N$, indicating that the phase transition at $T_N$ is of first order \cite{Ortega-San-Martin,Lee1}. The negative fitted values of ${\it \theta}$ mean the dominance of antiferrodistortive inter-structural-unit interactions which are stronger in MgCr$_2$O$_4$ (${\it \theta}$ = -14 K) than in ZnCr$_2$O$_4$ (${\it \theta}$ = -6 K). The dominance of such antiferrotype interactions in the PM phase seems to be compatible with the nonuniform lattice distortion in the AF phase revealed experimentally in ZnCr$_2$O$_4$, where neighboring tetrahedra are alternately stretched and compressed along the same crystal axis \cite{Ji}. In the antiferrotype nonuniform lattice order, the tetragonal lattice strain per spinel unit cell should be negligibly small, as is consistent with the very small magnitudes of the tetragonal lattice strain observed in $A$Cr$_2$O$_4$ in the AF phase \cite{Lee1,Ortega-San-Martin,Chung}.

\subsection{Softening with minimum in PM phase}
Next, we discuss the origin of the other salient elastic anomaly in the PM phase, the characteristic minimum at $\sim$50 K in $C_{44}(T)$. As shown in Fig. 3, this anomaly is observed in not only MgCr$_2$O$_4$ (open circles, from Fig. 1(c)) but also ZnCr$_2$O$_4$ (open squares, from Ref. [\cite{Kino}]). Taking into account the absence of orbital degeneracy for the $B$-site Cr$^{3+}$ in these compounds, the spin degrees of freedom should play a significant role for the presence of the anomaly in $C_{44}(T)$. One possible origin for the anomaly in $C_{44}(T)$ is the spin JT effect. However, the symmetry of the spin-lattice order in MgCr$_2$O$_4$ and ZnCr$_2$O$_4$ is tetragonal, which is different from the trigonal symmetry of $C_{44}$ \cite{Lee1,Ortega-San-Martin,Ji}. Thus, in the spin-JT-effect scenario, as described above, while the tetragonal shear modulus $C_t(T)$ should exhibit Curie-type softening in the cubic phase (PM phase) as expressed in Eq.~(\ref{eq:JT}), the trigonal shear modulus $C_{44}(T)$ should be anomaly-free as $C_{44}(T)=C_{44}^0$, which rules out the spin JT effect as a possible origin for the anomaly in $C_{44}(T)$. The other possible origin for the anomaly in $C_{44}(T)$ is the coupling between the dynamical molecular spin state and the acoustic phonons. We here note that the INS studies in MgCr$_2$O$_4$ as well as ZnCr$_2$O$_4$ observed the quasielastic magnetic scattering in the PM phase \cite{Lee1,Lee2,Suzuki,Tomiyasu}, and characterized this quasigapless mode as the fluctuations of the AF hexagonal spin clusters (the AF hexamers) in the pyrochlore lattice \cite{Lee2,Tomiyasu}. As illustrated in the inset to Fig. 3, the AF hexamers are aligned in the trigonal (111) planes of the spinel structure, and thus should be sensitive to the trigonal lattice deformation which generates the trigonal exchange striction. Therefore, {\it the anomaly of the trigonal shear modulus $C_{44}(T)$ in the PM phase should stem from the dynamical molecular spin state.}

The ultrasound velocity measurements in the present study extract information about the acoustic phonon dispersions at very small wave vectors of $q\sim 10^{-5}$ $\AA^{-1}$. In contrast, the molecular spin fluctuations observed in the INS spectra have large wave vectors of $q\sim1.5$ $\AA^{-1}$ \cite{Tomiyasu,Lee1}. Thus the observation of the anomaly in $C_{44}(T)$ suggests that the strong coupling of the dynamical molecular spin state to the lattice modifies the acoustic phonon dispersions over a wide range in the Brillouin zone, even at around zero wave vector. The modification of the phonon dispersions would be strongest at around $q\sim1.5$ $\AA^{-1}$.

\begin{figure}[t]
\begin{center}
\includegraphics[scale=0.4]{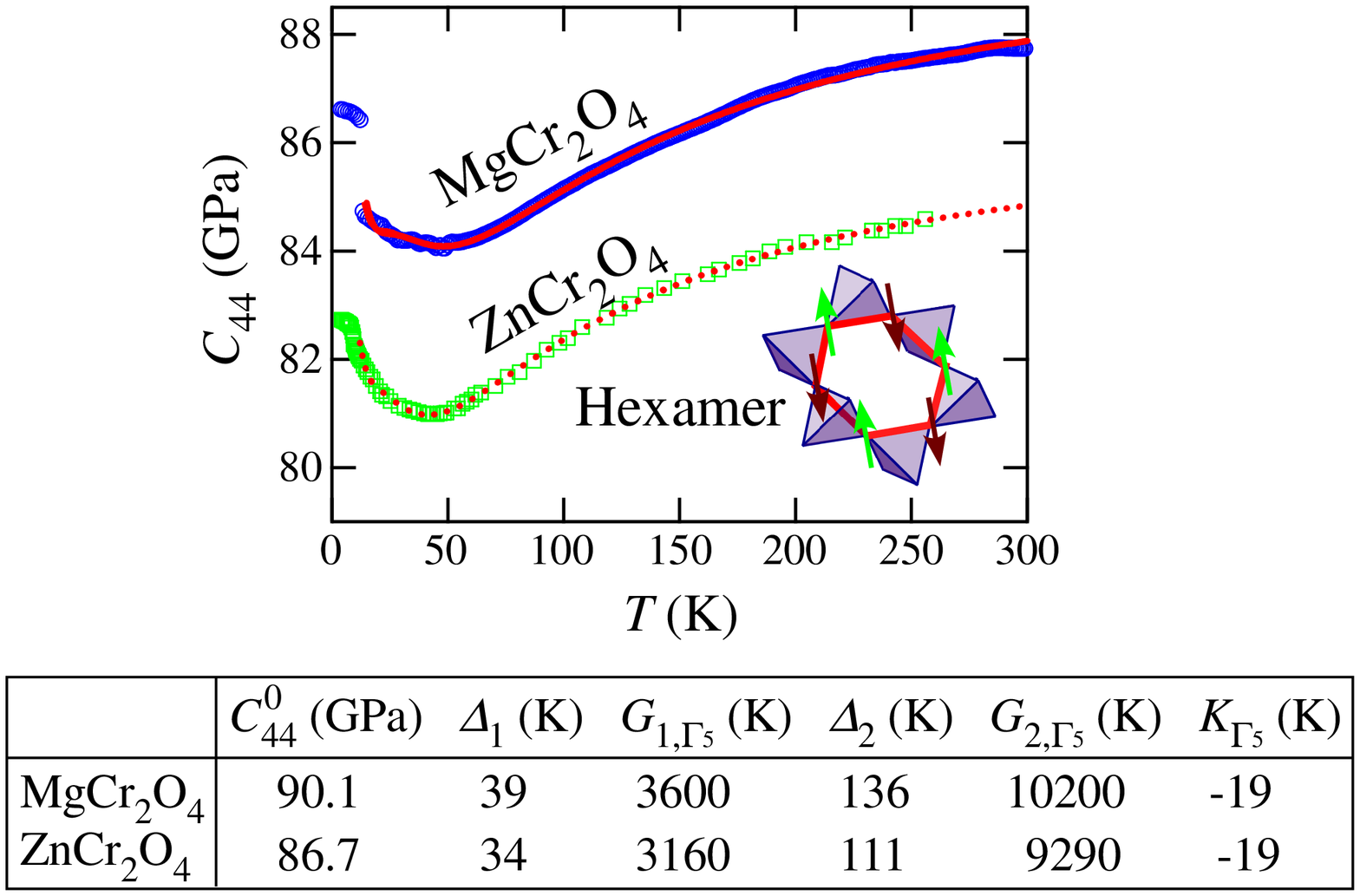}
\caption{\label{fig:Fig3} (Color online). The trigonal shear modulus $C_{44}$ in MgCr$_2$O$_4$ (open circles, from Fig. 1(c)) and ZnCr$_2$O$_4$ (open squares, from Ref. [\cite{Kino}]) as functions of temperature. The solid and dotted curves are fits of the experimental data for MgCr$_2$O$_4$ and ZnCr$_2$O$_4$ to Eq.~(\ref{eq:SM}), respectively. Values for the fitting parameters are listed in the table beneath the figure. The inset depicts an AF hexamer in the Cr$^{3+}$ pyrochlore lattice.}
\end{center}
\end{figure}

Regarding the elastic anomaly from the short-range spin correlations, the characteristic minimum in $C_{44}(T)$ in MgCr$_2$O$_4$ and ZnCr$_2$O$_4$ is similar to that in low-dimensional spin-dimer systems such as SrCu$_2$(BO$_3$)$_2$ \cite{Wolf}. Notably, the minimum in $C_{\Gamma}(T)$ in the spin dimer systems is explained as a result of the presence of a finite gap for the local magnetic excitations which is sensitive to the strain \cite{Wolf}. In the spin-dimer systems, the quantitative analysis of the experimental data of $C_{\Gamma}(T)$ is given by applying a theoretical model which has the same form as Eq.~(\ref{eq:SS}). Assuming a singlet-triplet excitation gap $\Delta_1$ for a single spin dimer and a multitriplet excitation gap $\Delta_2$, the contribution of the spin dimers to the elastic constant takes the form, \cite{Wolf}
\begin{equation}
C_{\Gamma}(T)=C_{\Gamma}^0-G_{1,\Gamma}^2N\frac{\chi_{\Gamma}(T)}{(1-K_{\Gamma}\chi_{\Gamma}(T))},
\label{eq:SM}
\end{equation}
with $C_{\Gamma}^0$ the background elastic constant, $N$ the density of spin dimers, $G_{1,\Gamma}=|\partial \Delta_1/\partial \epsilon_{\Gamma}|$ the coupling constant for a single spin dimer measuring the strain ($\epsilon_{\Gamma}$) dependence of the excitation gap $\Delta_1$, $K_{\Gamma}$ the ($q$ = 0) dimer-dimer interaction, and $\chi_{\Gamma}(T)$ the strain susceptibility of a single spin dimer. Eq.~(\ref{eq:SM}) is based on the coupling between the gapped spin triplet states and the acoustic phonons, which, in turn, means that this equation can be applied to all systems with the coupling between gapped states and acoustic phonons regardless of whether it is a singlet system or not. According to Eq.~(\ref{eq:SM}), the minimum in $C_{\Gamma}(T)$ appears when this elastic mode strongly couples to the excited state at $\Delta_i$; on cooling, $C_{\Gamma}(T)$ exhibits softening roughly down to $T\sim\Delta_i$, but recovery of the elasticity (hardening) roughly below $T\sim\Delta_i$.

In analogy to the spin dimer systems in the presence of local excitations, therefore, the minimum in $C_{44}(T)$ in MgCr$_2$O$_4$ and ZnCr$_2$O$_4$ should arise from a gap in the molecular spin excitations which is sensitive to the strain. This interpretation helps to understand the INS results for MgCr$_2$O$_4$ and ZnCr$_2$O$_4$. The broad quasielastic magnetic scattering spectrum in the PM phase comprises gapless spin fluctuations which probably form embryo of the tetragonal spin-lattice order below $T_N$, and gapped spin excitations which are considerably smeared \cite{Lee1,Lee2,Suzuki,Tomiyasu}. And the observation of distinct gapped excitations in the AF phase is attributed to the suppression of spin fluctuations by spin JT distortion \cite{Lee1,Suzuki,Tomiyasu}.

Similar to the spin dimer systems \cite{Wolf}, we now give a quantitative analysis of the experimental data of $C_{44}(T)$ in MgCr$_2$O$_4$ and ZnCr$_2$O$_4$ assuming an excitation gap $\Delta_1$ for a single AF hexamer and a multi-AF-hexamer excitation gap $\Delta_2$. The contribution of the AF hexamers to the elastic constant should take the same form as Eq.~(\ref{eq:SM}) with $C_{44}^0$ the background elastic constant, $N$ the density of AF hexamers, $G_{1,\Gamma_5}=|\partial \Delta_1/\partial \epsilon_{\Gamma_5}|$ the coupling constant for a single AF hexamer measuring the strain ($\epsilon_{\Gamma_5}$) dependence of the excitation gap $\Delta_1$, $K_{\Gamma_5}$ the ($q$ = 0) inter-AF-hexamer interaction, and $\chi_{\Gamma_5}(T)$ the strain susceptibility of a single AF hexamer. Here, we ignore the $T$-dependence in the background $C_{44}^0$ (the hardening with decreasing $T$) \cite{Varshni}, because the nonmonotonic variation in $C_{44}(T)$ is very large compared with the hardening on cooling in $C_{44}^0$. Fits of the experimental data for MgCr$_2$O$_4$ and ZnCr$_2$O$_4$ to Eq.~(\ref{eq:SM}) are depicted in Fig. 3 as solid and dotted curves, respectively. Here, the value of $N$ = 3.45$\times$10$^{27}$ m$^{-3}$ is fixed in a first approximation \cite{Lee2}. With the fitted parameters listed in the table beneath the figure, fits of Eq.~(\ref{eq:SM}) are in excellent agreement with the experimental data, reproducing the characteristic minimum in $C_{44}(T)$. For MgCr$_2$O$_4$ (ZnCr$_2$O$_4$), the fitted gap value of $\Delta_1$ = 39 K (34 K) in the PM phase is rather smaller than the gap of $\Delta\simeq$ 50 K measured in the AF phase \cite{Lee1,Tomiyasu}, which implies that the suppression of spin fluctuations by the spin JT distortion makes the gap in the AF phase, $\Delta$, rather larger than that in the PM phase, $\Delta_1$. The negative value of $K_{\Gamma_5}$ = -19 K means that the inter-AF-hexamer interaction is antiferrodistortive. The coupling constant $G_{2,\Gamma_5}=|\partial \Delta_2/\partial \epsilon_{\Gamma_5}|$ = 10200 K (9290 K) is about three times larger than $G_{1,\Gamma_5}$ = 3600 K (3160 K), indicating that the higher excitations $\Delta_2$ = 136 K (111 K) couple to the trigonal lattice deformation more strongly than the lowest excitations $\Delta_1$.

According to Eq.~(\ref{eq:ES2}), not only $C_{44}$ but also $C_{11}$ should couple to the AF hexamers via the exchange striction mechanism, while $C_t$ should be inactive. However, the nonmonotonic $T$ dependence is invisible in $C_{11}(T)$. Instead, $C_{11}(T)$ exhibit Curie-type softening due to the dynamical spin JT effect. We notice here that the magnitude of the nonomonotonic $T$ dependence in $C_{44}(T)$, especially the relative change in hardening on cooling below $\sim50$ K down to $T_N$ in $C_{44}(T)$ ($\Delta C_{44}/C_{44}\sim1\%$), is rather smaller than the magnitude of Curie-type softening in $C_{11}(T)$ ($\Delta C_{11}/C_{11}\sim10\%$). Thus, in $C_{11}(T)$, the weak nonmonotonic $T$ dependence would be hidden behind the huge Curie-type softening.

\subsection{Hardening in AF phase}
As described above, the elastic anomalies in MgCr$_2$O$_4$ and ZnCr$_2$O$_4$ in the PM phase strongly suggest {\it the coexistence of dynamical spin JT effect and dynamical molecular spin state, both of which strongly couple to the acoustic phonons.} Conversely, in the AF phase, $C_t(T)$ and $C_{44}(T)$ exhibit the usual hardening on cooling without anomaly, as shown in Figs. 2 and 3, respectively \cite{Varshni}. According to the INS experiments, while the spin JT distortion below $T_N$ partially releases frustration, the dynamical molecular spin state persists even in the AF phase indicating the survival of some frustration \cite{Tomiyasu,Lee1}. Thus, combined with the above discussion on the elastic anomalies in the PM phase, we attribute the absence of anomalies in $C_t(T)$ and $C_{44}(T)$ in the AF phase to different origins: respectively, the spin JT distortion (generating the partial release of frustration) for $C_t(T)$, and the decoupling of the acoustic phonons from the dynamical molecular spin state (producing the decoupling from the remaining frustration) for $C_{44}(T)$.

\section{Summary}
To summarize, ultrasound velocity measurements of MgCr$_2$O$_4$ revealed elastic-mode-dependent sound-velocity anomalies in the PM phase. These elastic anomalies can be attributed to a coexistence of the dynamical spin JT effect and the dynamical molecular spin state in the PM phase, the presence of which is compatible with the coexistence of the magnetostructural order and the dynamical molecular spin state in the AF phase. Further experimental and theoretical work is necessary to clarify the coexistence mechanism for the two different types of frustration effects in the geometrically-frustrated chromite spinels.

\section{Acknowledgments}
This work was partly supported by Grants-in-Aid for Young Scientists (B) (21740266) and Priority Areas (22014001) from MEXT of Japan.

\end{document}